\documentclass[journal]{IEEEtran}
\usepackage{amsmath}
\usepackage[pdftex]{graphicx}
\usepackage{algorithm}
\usepackage{algpseudocode}
\usepackage{leftindex}
\usepackage{amssymb}
\usepackage{subcaption}
\usepackage{graphicx}
\usepackage{amsthm}
\usepackage[justification=centering]{caption}

\providecommand{\keywords}[1]
{
  \small	
  \textbf{\textit{Keywords---}} #1
}

\theoremstyle{definition}

\theoremstyle{remark}
\newtheorem*{remark}{Remark}

\title{Optimal Observer Design Using Reinforcement Learning and Quadratic Neural Networks}
\author{Soroush Asri, Luis Rodrigues\\Department of Electrical Engineering, Concordia University}

\begin{document}

\maketitle

\begin{abstract}    
This paper introduces an innovative approach based on policy iteration (PI), a reinforcement learning (RL) algorithm, to obtain an optimal observer with a quadratic cost function. This observer is designed for systems with a given linearized model and a stabilizing Luenberger observer gain.
We utilize two-layer quadratic neural networks (QNN) for policy evaluation and derive a linear correction term using the input and output data.
This correction term effectively rectifies inaccuracies introduced by the linearized model employed within the observer design.
A unique feature of the proposed methodology is that the QNN is trained through convex optimization.
The main advantage is that the QNN’s input-output mapping has an analytical expression as a quadratic form, which can then be used to obtain a linear correction term policy. This is in stark contrast to the available techniques in the literature that must train a second neural network to obtain policy improvement.
It is proven that the obtained linear correction term is optimal for linear systems, as both the value function and the QNN's input-output mapping are quadratic.
The proposed method is applied to a simple pendulum, demonstrating an enhanced correction term policy compared to relying solely on the linearized model. This shows its promise for addressing nonlinear systems.

\end{abstract}


 \section{Introduction}
 
 Many control system methods assume that the system's states are available. However, in practice, the states may not be all measured, necessitating the design of a state observer.
Luenberger proposed an observer where the output error is multiplied by a gain to form  the correction term~\cite{luenberger_observer}. 
The Kalman filter, an optimal filter that minimizes the state estimation error variance, was proposed in reference~\cite{KForg} and it was proven that the steady state Kalman gain can be obtained by solving an algebraic Riccati equation. A disadvantage of both the methods in references~\cite{luenberger_observer}\cite{KForg} is that an accurate linear model of the system is needed to estimate the states. However, it's important to note that a linear model might not be able to accurately model the dynamics of certain nonlinear systems. To design optimal observers and controllers for complex nonlinear systems, many researchers have studied data-driven methods such as RL~\cite{sutton2018reinforcement}\cite{lewisoptimalbook}\cite{busoniu2017reinforcement}. RL has demonstrated exceptional performance in system control~\cite{benchmarkingRL}. 
The use of RL is motivated by key factors such as adaptability, handling complex dynamics, learning from real-world experience, and mitigating inaccuracies in system models~\cite{bertsekas2019reinforcement}\cite{kiumarsi2017optimal}.
Most research on using RL to design optimal controllers and observers is focused on discrete-time value-based RL algorithms~\cite{lewis2009reinforcement}. In value-based RL, the core concept is to estimate value functions, which are used to obtain the optimal policy. A common method to estimate value functions  is using the temporal difference (TD) equation~\cite{TD2018}. This equation reduces the Bellman error~\cite{tedrake2016underactuated}  by repeatedly updating the value function estimate. Value-based algorithms that use TD equation are called adaptive dynamic programming (ADP)~\cite{lewis2009reinforcement}. 
In control system, ADP is popular because it's closely related to dynamic programming and the Bellman equation~\cite{bertsekas2012dynamic}. The applications of ADP methods, the policy-iteration (PI) and the value-iteration (VI) algorithms, to feedback control are discussed in references~\cite{lewis2009reinforcement}\cite{prokhorov1997adaptive}. 
In the context of optimal control, PI is often preferred over VI due to its stability. PI guarantees policy improvement, which ensures reliable convergence to the optimal policy~\cite{LQR_RL_stable}.

In general, standard neural networks (SNN) are employed to approximate the value function during the policy evaluation step.
This approach is widely utilized for the development of a versatile ADP algorithm, which can effectively design many optimal controllers and observers~\cite{khan2012reinforcement}\cite{ADHDP}.
However, the utilization of SNNs as a value function approximator (VFA) has certain limitations:
    (i) the optimization of the neural network's weights is a non-convex problem. Consequently, training the neural network typically yields locally optimal weight,
    (ii) the process of selecting the most appropriate architecture for the SNN often involves a trial and error procedure, 
    (iii) the SNN lacks a straightforward analytical expression. Therefore, to improve the policy a second neural network is needed to minimize the value function,
    (iv) providing a comprehensive proof of convergence to the optimal policy is difficult or unattainable.
To address the mentioned issues, a two-layer quadratic neural network (QNN), as detailed in reference~\cite{QNN}, can be selected as the VFA. The advantages of using the two-layer QNN as the VFA over non-quadratic neural networks are:
    (i) two-layer QNNs are trained by solving a convex optimization, ensuring the globally optimal weights are obtained~\cite{QNN},
    (ii) the convex optimization also yields the optimal number of neurons in the hidden layer~\cite{Luis_QNN},
    (iii) the input-output mapping of the QNN is a quadratic form~\cite{Luis_QNN}. This allows analytical minimization of the value function with respect to the policy.
Reference~\cite{Luis_QNN} demonstrates the effectiveness of QNNs in various applications, including regression, classification, system identification, and the control of dynamic systems.
Furthermore, the utility of employing a two-layer QNN as the VFA becomes evident when addressing linear quadratic (LQ)~\cite{tedrake2016underactuated} problems, as both the value function and the input-output mapping of the QNN have a quadratic form. Also, the closed-form solution of the LQ problem and its relevance in various engineering applications have made it a popular criteria for comparing RL algorithms~\cite{matni2019self}. Therefore, PI algorithms have been applied to solve linear quadratic regulator (LQR), linear quadratic Gaussian (LQG), and linear quadratic tracker (LQT) problems, as discussed in references~\cite{bradtke1994adaptive}\cite{lewis2010reinforcement}\cite{kiumarsi2014reinforcement}, respectively. 
References~\cite{na2017adaptive}\cite{li2020networked} use the PI algorithm to obtain the optimal observer gain for LTI systems, where the performance index is chosen as a quadratic function with respect to the correction term and the output error.

 The objective of this paper is to design the optimal observer with a local cost quadratic in the output error and the correction term, employing QNNs as the VFA within the PI algorithm. It is based on the assumption that a linearized  model and a stabilizing Luenberger observer is provided. This paper proves that the porposed approach converges to the optimal observer obtained from closed-form solutions, if the linear model is accurate. The proposed method is also applied to a simple pendulum, which gives an improved observer compared to relying solely on the linearized model, highlighting its potential for nonlinear system applications.
To the best of our knowledge, this is the first time that the optimal observer is obtained with QNNs trained by solving a convex optimization. 
This paper is organized as follows. Section~\ref{sec: Problem Statement for LQE} presents the problem statement and section~\ref{sec: general Policy iteration} gives the PI algorithm.
After section~\ref{section: QNN} reviews the two-layer QNN, section~\ref{section: proposed alg} presents how to perform the PI algorithm using the two-layer QNN. Section~\ref{sec: simulation} shows  the simulation results and section~\ref{sec: conclusion} concludes the paper. 

\section{Problem Statement} 
\label{sec: Problem Statement for LQE}

  Consider the provided linearized model as
\begin{gather}
    \nonumber x_{k+1} = Ax_k + Bu_k \\
    y_k = C x_k \label{eq: system model } 
\end{gather}

\noindent where $x_k \in 	\mathbb{R}^{n_x}$ is the unknown state vector, $u_k \in 	\mathbb{R}^{n_u}$ is the input vector, $y_k \in \mathbb{R}^{n_y}$ is the measured output vector, and $A$, $B$, $C$ are system matrices. Assume $(A,C)$ is observable. The goal is to design an optimal observer for the system and minimize a performance index.
The dynamics of the observer utilizing the provided model are as follows
\begin{gather}
    \nonumber \hat{ x }_{k+1} = A \hat{x}_k  + Bu_k + w_k^\pi  \\
    \hat{y}_k = C \hat{x}_k \label{eq: Luenberger observer } 
\end{gather}
where $\hat{x}_{k}$ and $\hat{y}_{k}$  are states and outputs estimates, respectively, and $w_k^\pi = \pi(.) $ is the correction term policy to be designed later.
The value-function following policy $\pi(.)$ is written as
\begin{gather}
    \nonumber
    V^\pi(\Tilde{x}_k) = \sum_{i=k}^\infty \gamma^{i-k} \left ( \Tilde{y}_i^T Q \Tilde{y}_i +   \left (w_i^\pi  \right )^T R w_i^\pi \right )  = \sum_{i=k}^\infty \gamma^{i-k} c(\Tilde{y}_i, w_i^\pi)  
    \label{eq: performance index of LQE}
\end{gather}
where  $Q = Q^T \geq 0 $, $R = R^T > 0 $  are chosen,  $c(\Tilde{y}_k, w_k^\pi)   $ is the local cost at time step $k$, $ 0 \leq \gamma < 1$ is the discount factor,  $\Tilde{y}_k = y_k - \hat{y}_k$, and $\Tilde{x}_{k} = {x}_{k} - \hat{x}_{k}$.  The objective is to design the correction term policy $w_k^\pi$ that minimizes the value-function using PI algorithm.

The Bellman equation~\cite{bertsekas2019reinforcement} is
\begin{gather}
V^\pi(\Tilde{x}_{k}) = \Tilde{y}_k^T Q \Tilde{y}_k  +  (w_k^\pi)^T R w_k^\pi   + \gamma V^\pi (\Tilde{x}_{k+1}) 
    \label{eq: Bellman of LQE}
\end{gather}
  In order to use the PI algorithm, we must add the assumption that an initial policy $\pi_0(.)$ which stabilizes the observer error dynamics is known.  The general PI algorithm to obtain the optimal policy $\pi^*(.)$ is given in Algorithm~\ref{alg: PI for LQE}.




     \begin{algorithm}[H]
     \small
        \caption{General Policy Iteration } \label{alg: PI for LQE}
          
        \begin{algorithmic} 
            \State Select the initial policy $\pi_0$ that stabilizes the observer error dynamics. Then, for $j=0,1,2,\hdots$ perform policy evaluation and policy improvement steps until  convergence 
            \State \textbf{Policy Evaluation:} 
            \State  $\: \: \: $ Solve for $V^{\pi_{j}}(.)$ such that
            \begin{equation}
                \begin{split} 
                    V^{\pi_{j}}(\Tilde{x}_{k}) = \Tilde{y}_k^T Q \Tilde{y}_k  +  (w_k^{\pi_{j}})^T R w_k^{\pi_{j}}   + \gamma V^{\pi_{j}} (\Tilde{x}_{k+1}) 
                \end{split}
            \end{equation}

            \State \textbf{Policy improvement:}
            \State $\: \: \: \: \:$ Obtain $w_k^{\pi_{j+1}}$ such that
            \begin{equation}
                \begin{split} 
                    w_k^{\pi_{j+1}} =  \arg  \min_{{\pi_j}} V^{\pi_{j}}(\Tilde{x}_{k})
                \end{split}
            \end{equation}

        \end{algorithmic}
    \end{algorithm}

\begin{remark}
\label{remark: linear system}
When the system can be precisely described by the linear model, the error dynamics is
\begin{gather}
  \nonumber \Tilde{x}_{k+1} = A\Tilde{x}_{k}  -w_k^\pi \\
  \Tilde{y}_k = C \Tilde{x}_k \label{eq: error system}
\end{gather}
and the goal is to solve
\begin{equation} 
    \begin{split}
    \label{eq: objective}
        \arg \min_{\pi } \: \: \: &  \sum_{i=k}^\infty  \gamma^{i-k} \left (\Tilde{y}_i^T Q \Tilde{y}_i +  \left  (w_i^\pi  \right )^T R w_i^\pi   \right ), \: \: \forall k   \\
        s.t. \: \: \: & \Tilde{x}_{i+1} = A\Tilde{x}_{i} - w_i^\pi \\
                      &\Tilde{y}_i = C \Tilde{x}_i 
    \end{split}
    \end{equation}
    The optimization problem~\eqref{eq: objective} can be rewritten as
    \begin{equation} 
    \begin{split}
    \label{eq: objective chang to LQR}
        \arg \min_{\pi } \: \: \: &  \sum_{i=k}^\infty  \gamma^{i-k} \left (\Tilde{x}_i^T \Bar{Q} \Tilde{x}_i +   \left (w_i^\pi \right )^T R w_i^\pi \right ), \: \forall k   \\
        s.t. \: \: \: & \Tilde{x}_{i+1} = A\Tilde{x}_{i} + (-I_{n_x})w_i^\pi  
    \end{split}
    \end{equation}

\noindent where $\Bar{Q} = C^T Q C$, $I_{n_x}$ is the $ n_x\times n_x
$ identity matrix. Note that $(A,-I_{n_x})$ is controllable. Therefore, the optimization problem~\eqref{eq: objective chang to LQR} can be considered as an LQR problem for the observer error dynamics. Thus, the optimal policy $\pi^*(.)$ is a linear function, and the optimal value-function  $V^{\pi^*}(.)$ is quadratic, which makes the QNN a perfect candidate to approximate it. Therefore,
we  consider $\pi( \Tilde{x}_k) = K^\pi \Tilde{x}_k$ for some matrix $K^\pi$, and  the corresponding value-function approximator $V^{\pi}(\Tilde{x}_k) = \Tilde{x}_k^T P^\pi \Tilde{x}_k$ for a unique matrix $P^\pi>0$~\cite{lewisoptimalbook}.. 
\end{remark}
   \begin{remark}
       
    If $\Tilde{x}_k$ is known, one can perform algorithm~\ref{alg: PI for LQE} as presented in~\cite{bradtke1994adaptive}. However, $\Tilde{x}_k$ is unknown and
  one cannot perform the policy evaluation step. To address this problem, we write the Bellman equation in terms of previous measured data instead of $\Tilde{x}_k$ and $\Tilde{x}_{k+1}$ and revise the PI algorithm accordingly.
   \end{remark}
   





\section{Revised Policy Iteration Algorithm}
\label{sec: general Policy iteration}
In this section, we present the refined PI algorithm. First, we reformulate the Bellman equation by incorporating previous measurement data.
\subsection{Refining the Bellman Equation}  
Following the approach in reference~\cite{lewis2010reinforcement}, we reconstruct $\Tilde{x}_k$ by  previous measured data and replace $\Tilde{x}_k$ in the Bellman equation. Consider the observer error dynamics~\eqref{eq: error system} as the expanded state equation~\cite{lewis2010reinforcement}    
        \begin{equation}
            \begin{split}
            \label{eq: expanded state equation}
                 & \Tilde{x}_k = A^{n_x} \Tilde{x}_{k-n_x}  + \mathcal{C} w_{k-1,k-n_x}^\pi \\
                          & \Tilde{y}_{k-1,k-n_x} = \mathcal{O} \Tilde{x}_{k-n_x} + T  w_{k-1,k-n_x}^\pi
            \end{split}
        \end{equation}  
         where   
     \begin{gather}
         \nonumber w_{k-1,k-n_x}^\pi = \begin{bmatrix}
                            w_{k-1}^\pi \\ w_{k-2}^\pi \\ \vdots \\ w_{k-n_x}^\pi
                       \end{bmatrix}, \: \: \:
        \Tilde{y}_{k-1,k-n_x} = \begin{bmatrix}
                            \Tilde{y}_{k-1} \\ \Tilde{y}_{k-2} \\ \vdots \\ \Tilde{y}_{k-n_x}
                       \end{bmatrix},  \\
        \nonumber T = \begin{bmatrix}
                0  &  -C  &  -CA  &  \hdots  & -CA^{n_x -2} \\
                0  &   0  &   -C  &   \hdots  &  -CA^{n_x -3} \\
                \vdots  &  \vdots &  \ddots  & \ddots   & \vdots \\
                0  &  \hdots &  &  0 & -C  \\
                0 & 0 & 0 & 0 & 0
            \end{bmatrix},              
    \end{gather}  
    \noindent  $\mathcal{C} = 
    \begin{bmatrix}    
                -I_{n_x}  &  -A  &  -A^2  &  \hdots  &  -A^{n_x-1}                    
     \end{bmatrix}$, and the observability matrix $\mathcal{O}$ is 
    \begin{gather}
       \mathcal{O} = \begin{bmatrix}
            CA^{n_x-1} \\ CA^{n_x-2} \\ \vdots \\ CA \\ C
       \end{bmatrix} \in 	\mathbb{R}^{(n_xn_y) \times n_x }.
    \end{gather}
    Note that $(A,C)$ is observable, and the observability matrix $\mathcal{O}$ has full column rank $n_x$. Therefore, its left inverse $\mathcal{O}^+$ is obtained by $ (\mathcal{O}^T \mathcal{O})^{-1} \mathcal{O}^T$.
    The following Lemma based on reference~\cite{lewis2010reinforcement} allows us to replace $\Tilde{x}_k$ with measured data.
\vspace{5pt}
 
    \noindent \textbf{Lemma 1.}  One can reconstruct $\Tilde{x}_k$ from measured data as  
    \begin{gather}
        \Tilde{x}_k = \begin{bmatrix}
                        M_w & M_{\Tilde{y}}
                    \end{bmatrix} 
                    \begin{bmatrix}
                        w_{k-1,k-n_x}^\pi \\ \Tilde{y}_{k-1,k-n_x}
                    \end{bmatrix} \label{eq: reconstruct the state}
    \end{gather}
where $M_w = \mathcal{C} - A^{n_x}\mathcal{O}^+ T$ and $M_{\Tilde{y}} = A^{n_x} \mathcal{O}^+$. 
    \begin{proof}
    Equation~\eqref{eq: expanded state equation} yields
    \begin{equation}
        \begin{split}
             \Tilde{x}_k - \mathcal{C} w_{k-1,k-n_x}^\pi &=  A^{n_x} \Tilde{x}_{k-n_x}   \\
             & = A^{n_x}  I_{n_x} \Tilde{x}_{k-n_x}
        \end{split}
    \end{equation} 
    where $I_{n_x}$ is the $n_x \times n_x$ identity matrix. Note that $I_{n_x} = \mathcal{O}^+ \mathcal{O}$. Therefore,
    \begin{equation}
        \begin{split} \label{eq: argrgr}
             \Tilde{x}_k - \mathcal{C} w_{k-1,k-n_x}^\pi  =  A^{n_x} \mathcal{O}^+  (\mathcal{O} \Tilde{x}_{k-n_x}  )
        \end{split}
    \end{equation}  
    According to~\eqref{eq: expanded state equation}, we can replace $\mathcal{O} \Tilde{x}_{k-n_x} $ with $\Tilde{y}_{k-1,k-n_x} - T w_{k-1,k-n_x}^\pi$ in equation~\eqref{eq: argrgr} and get 
    \begin{equation}
        \begin{split} 
             \Tilde{x}_k - \mathcal{C} w_{k-1,k-n_x}^\pi  =  A^{n_x} \mathcal{O}^+  \left ( \Tilde{y}_{k-1,k-n_x} - T w_{k-1,k-n_x}^\pi \right ) 
        \end{split}
    \end{equation} 
which can be recast as 
    \begin{equation}
        \begin{split} 
             \Tilde{x}_k  & =  \begin{bmatrix}
                                    \mathcal{C}- A^{n_x} \mathcal{O}^+ T  &  A^{n_x} \mathcal{O}^+
                                \end{bmatrix}
                                \begin{bmatrix}
                                    w_{k-1,k-n_x}^\pi \\ \Tilde{y}_{k-1,k-n_x}
                                \end{bmatrix}    \\
                          & =
                    \begin{bmatrix}
                        M_w & M_{\Tilde{y}}
                    \end{bmatrix} 
                    \begin{bmatrix}
                        w_{k-1,k-n_x}^\pi \\ \Tilde{y}_{k-1,k-n_x}
                    \end{bmatrix} 
        \end{split}
    \end{equation}

        
        
        


        
    %

    \end{proof} 
Using Lemma 1, the Bellman equation can be written based on previous measured data as

\begin{multline}
                       \label{eq: quadratic cost-to-go for LQE}
                    \begin{bmatrix}
                        w_{k-1,k-n_x}^\pi \\ \Tilde{y}_{k-1,k-n_x}
                    \end{bmatrix}^T  H^\pi 
                    \begin{bmatrix}
                        w_{k-1,k-n_x}^\pi \\ \Tilde{y}_{k-1,k-n_x}
                    \end{bmatrix} = \Tilde{y}_k^T Q \Tilde{y}_k + (w_k^\pi)^T R w_k^\pi   + \\ \gamma
                     \begin{bmatrix}
                        w_{k,k-n_x+1}^\pi \\ \Tilde{y}_{k,k-n_x+1}
                    \end{bmatrix}^T  H^\pi
                    \begin{bmatrix}
                        w_{k,k-n_x+1}^\pi \\ \Tilde{y}_{k,k-n_x+1}
                    \end{bmatrix} 
\end{multline} 
where $H^\pi = \begin{bmatrix}
            M_w & M_{\Tilde{y}}
        \end{bmatrix}^T
        P^\pi \begin{bmatrix}
            M_w & M_{\Tilde{y}}
        \end{bmatrix} $  is a symmetric matrix.
One contribution of this paper is to use the available measurements to train a two-layer QNN with a single output to find the matrix $H^\pi$ of equation (\ref{eq: quadratic cost-to-go for LQE}) and evaluate the policy $\pi(.)$. An overview of two-layer QNNs is provided in section~\ref{section: QNN}.
The neural network training algorithm is discussed in section~\ref{section: proposed alg}.
        
        
        \vspace{5pt}
   \begin{remark}
          Note that~\eqref{eq: quadratic cost-to-go for LQE} is a scalar equation.
        \begin{gather}
            \begin{bmatrix}
                w_{k-1,k-n_x}^\pi \\ \Tilde{y}_{k-1,k-n_x}
            \end{bmatrix} \in \mathbb{R}^{n_x(n_x + n_y)}
        \end{gather}
and $H^\pi$ is symmetric. Therefore, the matrix $H^\pi$ has  $M = \frac{n_x \left(n_x + n_y \right) \left(n_x \left(n_x + n_y \right)+1 \right)}{2}$ unknown independent elements and $ N \geq M$  data samples are needed to obtain $H^\pi$ from~\eqref{eq: quadratic cost-to-go for LQE}. 
 \end{remark}    
     \begin{remark}
          PI algorithms require persistent excitation (PE)~\cite{lewis2009reinforcement},~\cite{lewis2010reinforcement}. To achieve PE in practice, we add a probing noise term $n_k$ such that $w_k^\pi = K^\pi \Tilde{x}_k + n_k$. 
It is shown in reference~\cite{lewis2010reinforcement} that the solution computed by PI differs from the actual value corresponding
to the Bellman equation when the probing noise term $n_k$ is added. It is discussed that adding the discount factor $0<\gamma<1$ to the Bellman equation  reduces this harmful effect of $n_k$.
\end{remark}

       \subsection{Policy improvement step} 
        
   We now address how to improve the policy $\pi(.)$ after evaluating the policy and obtaining the matrix $H^\pi$ with a two-layer QNN.
    The policy improvement step can be written as
    \begin{equation}
        \begin{split} 
        \pi^{\prime}(\Tilde{x}_k) = \arg  \min_\pi  ( & \Tilde{y}_k^TQ\Tilde{y}_k + (w_k^\pi)^T R w_k^\pi + \\ 
        & \gamma \begin{bmatrix}
            w_{k,k-n_x+1}^\pi \\ \Tilde{y}_{k,k-n_x+1}
        \end{bmatrix}^T  H^\pi
        \begin{bmatrix}
            w_{k,k-n_x+1}^\pi \\ \Tilde{y}_{k,k-n_x+1}
        \end{bmatrix}  )
        \label{eq:  policy improvement step}
        \end{split}
    \end{equation}

    \noindent where $\pi^{\prime}(.)$ is the improved policy over the policy $\pi(.)$.
    Partition $\begin{bmatrix}
                    w_{k,k-n_x+1}^\pi \\ \Tilde{y}_{k,k-n_x+1}
                \end{bmatrix}^T  H^\pi 
                \begin{bmatrix}
                    w_{k,k-n_x+1}^\pi \\ \Tilde{y}_{k,k-n_x+1}
                \end{bmatrix}$ as
    \begin{gather}
        \begin{bmatrix}
        w_k^\pi \\ w_{k-1,k-n_x+1}^\pi \\ \Tilde{y}_{k,k-n_x+1} 
        \end{bmatrix}^T
        \begin{bmatrix} H_{11}^\pi & H_w^\pi & H_{\Tilde{y}}^\pi \\ ({H_w^\pi})^T & H_{22}^\pi & H_{23}^\pi \\ {(H_{\Tilde{y}}^\pi)}^T & {(H_{23}^\pi)}^T & H_{33}^\pi  \end{bmatrix}
        \begin{bmatrix}
            w_k^\pi \\ w_{k-1,k-n_x+1}^\pi \\ \Tilde{y}_{k,k-n_x+1} 
        \end{bmatrix}
    \end{gather}

    One can solve~\eqref{eq:  policy improvement step} and get the improved policy as
  \begin{gather}
        w_k^{\pi^\prime} = - \gamma (R + \gamma H_{11}^\pi)^{-1} (H_w^\pi w_{k-1,k-n_x+1} +  H_{\Tilde{y}}^\pi\Tilde{y}_{k,k-n_x+1})  
  \end{gather}



   Therefore, we can use the refined PI Algorithm~\ref{alg: PI_LQE_vanila} and obtain the optimal policy $\pi^*(.)$ without the system model.

       \begin{algorithm}
        \caption{Refined Policy Iteration} \label{alg: PI_LQE_vanila}
        \begin{algorithmic} 
        
            \State Select the initial policy $w_k^{\pi_0} $ that stabilizes the observer error dynamics. Then, for $j=0,1,2,\hdots$ perform policy evaluation and policy improvement steps until  convergence 
            \State \textbf{Policy Evaluation:} 
            \State  $\: \: \: \: \:$ Solve for $H^{\pi_{j}}$ using a two-layer QNN such that

            \begin{equation}
                \begin{split} 
                \label{eq: asdhfhfhj}
                \nonumber
                    \: \: \: \: \:  
                     & \begin{bmatrix}
                        w_{k-1,k-n_x}^{\pi_{j}} \\ \Tilde{y}_{k-1,k-n_x}
                    \end{bmatrix}^T  H^{\pi_{j}} 
                    \begin{bmatrix}
                        w_{k-1,k-n_x}^{\pi_{j}} \\ \Tilde{y}_{k-1,k-n_x}
                    \end{bmatrix}= \Tilde{y}_k^T Q \Tilde{y}_k + \\  &
                    (w_k^{\pi_{j}})^T R w_k^{\pi_{j}}   + \gamma \begin{bmatrix}
                        w_{k,k-n_x+1}^{\pi_{j}} \\ \Tilde{y}_{k,k-n_x+1}
                    \end{bmatrix}^T  H^{\pi_{j}} 
                    \begin{bmatrix}
                        w_{k,k-n_x+1}^{\pi_{j}} \\ \Tilde{y}_{k,k-n_x+1}
                    \end{bmatrix}  
                \end{split}
            \end{equation}

            \State \textbf{Policy improvement:}
            \State $\: \: \: \: \:$ Obtain ${\pi_{j+1}}$  such that
            
            \begin{equation}
                \begin{split} 
                    \nonumber
                    \: \: \: \: \: w_k^{\pi_{j+1}}  = -   & \gamma (R+ \gamma H_{11}^{\pi_{j}})^{-1} (H_w^{\pi_{j}} w_{k-1,k-n_x+1}   + \\ &     H_{\Tilde{y}}^{\pi_{j}} \Tilde{y}_{k,k-n_x+1}) 
                \end{split}
            \end{equation}

        \end{algorithmic}
    \end{algorithm}

     \begin{remark}
          For linear systems, the optimal $H^\pi$ can be obtained using the following closed-form solution
    \begin{gather}
            H^{\pi^*} = \begin{bmatrix}
            M_w & M_{\Tilde{y}}
        \end{bmatrix}^T
        P^{\pi^*} \begin{bmatrix}
            M_w & M_{\Tilde{y}}
        \end{bmatrix} 
    \end{gather}
 where $P^{\pi^*}$ is calculated from solving the Riccati equation~\eqref{eq: Riccati for LQE with discount factor}.~\cite{lewis2010reinforcement}
         \begin{multline}
            \label{eq: Riccati for LQE with discount factor}
              P^{\pi^*} = C^TQC + \gamma A^T P^{\pi^*} A   - \gamma^2 A^T P^{\pi^*} \\  (R+  \gamma P^{\pi^*}  )^{-1} P^{\pi^*} A 
         \end{multline}
    Then the optimal correction term can be designed as
            \begin{equation}
                \begin{split} 
                     w_k^{\pi^{*}}  = -   \gamma (R+ \gamma H_{11}^{\pi^{*}})^{-1} ( & H_w^{\pi^{*}} w_{k-1,k-n_x+1}   + \\  & H_{\Tilde{y}}^{\pi^{*}} \Tilde{y}_{k,k-n_x+1})   
                \end{split}
            \end{equation} 
            \end{remark}


    \section{Two-layer QNNs with one output}
\label{section: QNN}
This section discussed the training of a two-layer QNN  with one output as introduced in~\cite{QNN}. 
A QNN will be used to find the matrix $H^\pi$ of equation (\ref{eq: quadratic cost-to-go for LQE}) based on measured data.
Consider the neural network 
with one hidden layer, one output, and a degree two polynomial activation function, where  $X_i \in \mathbb{R}^n$ is the $i$-th input data given to the neural network, $\hat{Y}_i \in \mathbb{R} $ is the output of the neural network for the input $X_i$, $Y_i \in \mathbb{R}$ is the output label corresponding to  the input $X_i$, $L$ is the number of hidden neurons, $f(z)=az^2+bz+c$ is the polynomial activation function,  and $a \neq 0$, $b$, $c$ are pre-defined constant coefficients. The notation $w_{kj}$ represents the weight from the $k$-th input-neuron  to the $j$-th hidden-neuron, and $v_{j}$ represents the weight from the j-th hidden-neuron to the output.
The input-output mapping of the neural network is
      
      \begin{equation}
          \hat{Y}_i=\sum_{j=1}^{L} f(X_i^TW_j)v_{j} 
      \end{equation}

      \noindent where $W_j = \begin{bmatrix}
        w_{1j} &
        w_{2j} &
        \hdots &
        w_{nj}
      \end{bmatrix}^T$.      
Reference~\cite{QNN} proposes the training optimization 
\begin{equation}
    \begin{split} \label{eq: changed neural net optimizaion}
      \min_{W_k,v_k} \: \: & l(\hat{Y}-Y) + \beta \sum_{j=1}^{L} | v_j |\\
        s.t. \: \: & \hat{Y}_i=\sum_{j=1}^{L} f(X_i^TW_j)v_{j}\\
         & \Vert  W_k \Vert_2 =1, \: \: \: \: k=1,2,...,L \: \: \: \:i=1,2,...,N 
    \end{split}
\end{equation}




\noindent where $\beta\ge 0$ is a pre-defined regularization parameter, $l(.)$ is a convex loss function, N is the number of data points, $\hat{Y} = \begin{bmatrix}
        \hat{Y}_1 &
        \hat{Y}_2 &
        \hdots &
        \hat{Y}_N
      \end{bmatrix}^T$, and $Y = \begin{bmatrix}
        Y_1 &
        Y_2 &
        \hdots &
        Y_N
      \end{bmatrix}^T$. 
The optimization problem~\eqref{eq: changed neural net optimizaion} can be equivalently solved by the dual convex optimization~\eqref{convex_formula} 
\begin{equation}
\begin{split} \label{convex_formula}
      \min_{Z^+ , Z^-} \: \: & l(\hat{Y}-Y) + \beta  \:  Trace(Z_{1}^+ + Z_{1}^-)  \\
        s.t. \: \: & \hat{Y}_{i}=aX_{i}^T(Z_{1}^+ - Z_{1}^-)X_{i}   +  bX_{i}^T(Z_{2}^+ - Z_{2}^-) + \\
        & \: \: \: \: \: \: \: \:  \: \:  c \:  Trace(Z_{1}^+ - Z_{1}^-), \\ 
          & Z^+=\begin{bmatrix} Z_{1}^+ &    Z_{2}^+ \\(Z_{2}^+)^T &Trace(Z_{1}^+) \end{bmatrix} \geq 0 , \\
          & Z^-=\begin{bmatrix} Z_{1}^- &    Z_{2}^- \\(Z_{2}^-)^T &Trace(Z_{1}^-) \end{bmatrix} \geq 0, \\
          &  i=1, 2, \hdots N 
\end{split}
\end{equation}     
where $Z_1^+$, $Z_2^+$, $Z_1^-$, $Z_2^-$ are optimization parameters~\cite{QNN}.
After training the neural network and obtaining $Z^+$, $Z^-$ from~\eqref{convex_formula}, the quadratic input-output mapping is 
    
    \begin{equation} \label{quadratic_mapping_QNN}
    \hat{Y}_i=\begin{bmatrix}
    X_i  \\ 1 \end{bmatrix}^T
    H
    \begin{bmatrix}
    X_i  \\ 1 \end{bmatrix} 
     \end{equation}
where
\begin{eqnarray*}
H=\begin{bmatrix} a(Z_{1}^+ - Z_{1}^-) &    0.5b(Z_{2}^+ - Z_{2}^-)  \\0.5b(Z_{2}^+ - Z_{2}^-)^T &
    c   \left [ Trace  \left (Z_{1}^+ - Z_{1}^-  \right ) \right ]
    \end{bmatrix}
\end{eqnarray*}
    
    \vspace{2mm}
    
    
   \begin{remark}
        If $b=c=0, \: a =1$ are chosen, the input-output mapping is $
    \hat{Y}_i=
    X^T_i  
    H
    X_i  $,  
    where $H=(Z_{1}^+ - Z_{1}^-) $. 
  \end{remark}

    \section{QNNs as the policy evaluator}
    \label{section: proposed alg}
    
     This section shows how a two-layer QNN executes the policy evaluation step and obtains $H^\pi$. Then, the complete algorithm to find $\pi^* (.)$ without the system model is presented. First, we introduce the following Lemma.
    
    \vspace{3mm}
    
    \noindent \textbf{Lemma 2: }
    Let $\hat{H}^\pi_i$ denote the $i$-th approximation of $H^\pi$ for the policy $\pi(.)$ that stabilizes the observer error dynamics.
    Starting with $i=1$ and any symmetric $\hat{H}^\pi_0$, iterating through equation~\eqref{eq: iterative Bellman_error} will result in $\hat{H}^\pi_i$ converging to $H^\pi$ provided $0\le\gamma<1$.
 \begin{multline}
            \label{eq: iterative Bellman_error}
            \begin{bmatrix}
            w_{k-1,k-n_x}^\pi \\ \Tilde{y}_{k-1,k-n_x}
        \end{bmatrix}^T  \hat{H}^\pi_i  
        \begin{bmatrix}
            w_{k-1,k-n_x}^\pi \\ \Tilde{y}_{k-1,k-n_x}
        \end{bmatrix} =  \Tilde{y}_k^T Q \Tilde{y}_k + (w_k^\pi)^T R w_k^\pi   +         \\
            \gamma 
            \begin{bmatrix}
            w_{k,k-n_x+1}^\pi \\ \Tilde{y}_{k,k-n_x+1}
        \end{bmatrix}^T  \hat{H}^\pi_{i-1}
        \begin{bmatrix}
            w_{k,k-n_x+1}^\pi \\ \Tilde{y}_{k,k-n_x+1}
        \end{bmatrix}
    \end{multline}


    \begin{proof}

    \noindent Applying the equation~\eqref{eq: iterative Bellman_error} recursively yields
\begin{multline*}
    \begin{bmatrix}
            w_{k-1,k-n_x}^\pi \\ \Tilde{y}_{k-1,k-n_x}
        \end{bmatrix}^T  \hat{H}^\pi_i  
        \begin{bmatrix}
            w_{k-1,k-n_x}^\pi \\ \Tilde{y}_{k-1,k-n_x}
        \end{bmatrix}=
    \sum_{j=0}^{i-1}   \gamma^j c(\Tilde{y}_{k+j},w_{k+j}^\pi) \\ 
              +\gamma^i \begin{bmatrix}
                      w_{k+i-1,k+i-n_x}^\pi  \\  \Tilde{y}_{k+i-1,k+i-n_x} 
              \end{bmatrix}  ^T  
              \hat{H}^\pi_0
              \begin{bmatrix}
                      w_{k+i-1,k+i-n_x}^\pi  \\  \Tilde{y}_{k+i-1,k+i-n_x} 
              \end{bmatrix}  
\end{multline*}
where $c(\Tilde{y}_k,w_k^\pi)$ was defined in section (\ref{eq: performance index of LQE}). Let $i \rightarrow \infty$. Then,    
      \begin{multline}
              \label{eq: 21313}
               \begin{bmatrix}
                  w_{k-1,k-n_x}^\pi \\ \Tilde{y}_{k-1,k-n_x}
              \end{bmatrix}^T  \hat{H}^\pi_\infty 
              \begin{bmatrix}
                  w_{k-1,k-n_x}^\pi \\ \Tilde{y}_{k-1,k-n_x}
              \end{bmatrix} =  \sum_{j=0}^{\infty} \gamma^j  c(\Tilde{y}_{k+j},w_{k+j}^\pi)  \\
             + \lim_{i \to \infty} \gamma^i
              \begin{bmatrix}
                      w_{k+i-1,k+i-n_x}^\pi  \\  \Tilde{y}_{k+i-1,k+i-n_x} 
              \end{bmatrix}  ^T  
              \hat{H}^\pi_0 
              \begin{bmatrix}
                      w_{k+i-1,k+i-n_x}^\pi  \\  \Tilde{y}_{k+i-1,k+i-n_x} 
              \end{bmatrix} 
      \end{multline}
Since $\lim_{i \to \infty} \gamma^i = 0$, for any $\hat{H}^\pi_0$ we also get     
       \begin{multline}
        \label{eq: show limit is zero}
            \lim_{i \to \infty}
            \gamma^i
              \begin{bmatrix}
                      w_{k+i-1,k+i-n_x}^\pi  \\  \Tilde{y}_{k+i-1,k+i-n_x} 
              \end{bmatrix}  ^T  
              \hat{H}^\pi_0 
              \begin{bmatrix}
                      w_{k+i-1,k+i-n_x}^\pi  \\  \Tilde{y}_{k+i-1,k+i-n_x} 
              \end{bmatrix} = 0
       \end{multline}
      Replacing~\eqref{eq: show limit is zero} in~\eqref{eq: 21313} yields 
    \begin{gather}
        \label{eq: usss}
                     \begin{bmatrix}
                  w_{k-1,k-n_x}^\pi \\ \Tilde{y}_{k-1,k-n_x}
              \end{bmatrix}^T  \hat{H}^{\pi}_\infty 
              \begin{bmatrix}
                  w_{k-1,k-n_x}^\pi \\ \Tilde{y}_{k-1,k-n_x}
              \end{bmatrix} =  \sum_{j=0}^{\infty}  \gamma^j c(\Tilde{y}_{k+j},w_{k+j}^\pi)
      \end{gather}        
       According to the Bellman equation~\eqref{eq: quadratic cost-to-go for LQE}, we have      
    \begin{gather}
        \label{eq: usss2}
                     \begin{bmatrix}
                  w_{k-1,k-n_x}^\pi \\ \Tilde{y}_{k-1,k-n_x}
              \end{bmatrix}^T  H^{\pi} 
              \begin{bmatrix}
                  w_{k-1,k-n_x}^\pi \\ \Tilde{y}_{k-1,k-n_x}
              \end{bmatrix} =  \sum_{j=0}^{\infty} \gamma^j c(\Tilde{y}_{k+j},w_{k+j}^\pi)
      \end{gather}  
       As a result of~\eqref{eq: usss}\eqref{eq: usss2}, $\hat{H}_{i}^{\pi}$ converges to $H^{\pi}$.
\end{proof}

Therefore, in the policy evaluation step, we can calculate $H^{\pi}$ if we can obtain $\hat{H}^{\pi}_i$ in equation~\eqref{eq: iterative Bellman_error} from the previous known $\hat{H}^{\pi}_{i-1}$ in the $i$-th iteration. We use a two-layer QNN to obtain $\hat{H}^{\pi}_i$.

\subsection{Obtaining $\hat{H}^{\pi}_i$ using a two-layer QNN}

Define

\vspace{2mm}

\begin{figure}[t]
\begin{center}
\includegraphics[width=3in]{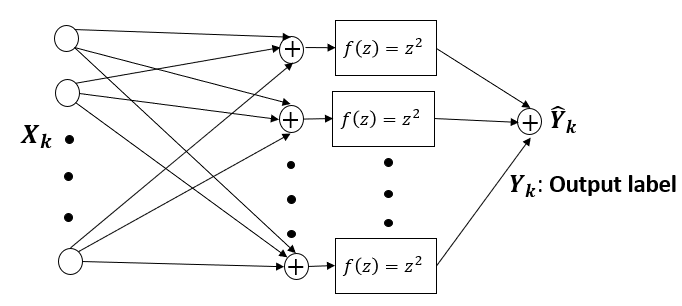}
\caption{QNN as the policy evalautor to obtain \\ $\hat{H}^{\pi}_i$ in i-th iteration }
\label{fig: QNN_in_LQE}
\end{center}
\end{figure}

            $X_k = \begin{bmatrix}
        w_{k-1,k-n_x}^\pi \\ \Tilde{y}_{k-1,k-n_x}
    \end{bmatrix}$

\begin{multline}
    Y_{k} = \Tilde{y}_k^T Q \Tilde{y}_k + (w_k^\pi)^T R w_k^\pi   + \\  \gamma  
        \begin{bmatrix}
            w_{k,k-n_x+1}^\pi \\ \Tilde{y}_{k,k-n_x+1}
        \end{bmatrix}^T  \hat{H}^\pi_{i-1}
        \begin{bmatrix}
            w_{k,k-n_x+1}^\pi \\ \Tilde{y}_{k,k-n_x+1}
        \end{bmatrix}
\end{multline}
Given $N >> M$ data points $X_k$ as the inputs and the corresponding data points $Y_{k}$ as the output labels to the QNN given in Fig~\ref{fig: QNN_in_LQE}, we train the QNN and obtain from (\ref{quadratic_mapping_QNN}) the quadratic input-output mapping as $X_k^T H X_k = \hat{Y}_{k}$. After obtaining $H$ we simply set $\hat H^\pi_i=H$.
The detailed algorithm~\ref{alg: PI_LQE_iterative} uses a convergence stopping criterium of $ || \hat{H}^{\pi_j}_i - \hat{H}^{\pi_j}_{i-1} || < \epsilon$ for a small $\epsilon$. 
       \begin{algorithm} 
       \small
        \caption{The Complete PI algorithm with a QNN as the VFA} \label{alg: PI_LQE_iterative}
        \begin{algorithmic} 
            \State Choose $\epsilon$, $N$, $\gamma$.
            \State Select the initial policy $w_k^{\pi_0}$ that stabilizes the observer error dynamics. Then, for $j=0,1,2,\hdots$ perform policy evaluation and policy improvement steps until  convergence 
            \State \textbf{Policy Evaluation:} 

            \State $\: \: \: \: \:$ $i \gets 0$   
            \State $\: \: \: \: \:$ Choose a random $\hat{H}^{\pi_j}_0$
            \State $\: \: \: \: \:$ \textbf{repeat}  
             \State $\: \: \: \: \:$  $\: \: \: \: \:$ $i \gets i+1 $

            \State $\: \: \: \: \:$ $\: \: \: \: \: $  Train the QNN by $N$ data samples with $X_k$ as the 
            \State $\: \: \: \: \:$ $\: \: \: \: \: \: \: \: \: \: \: \: \: \: \: \: \: \: \: \:$ input  and $Y_{k}$ as the output label.
            \State $\: \: \: \: \:$ $\: \: \: \: \:$ Obtain the input-output mapping as $ X_k^T H X_k = \hat{Y}_{k}$
            \State $\: \: \: \: \:$ $\: \: \: \: \:$ $\hat{H}^{\pi_{j}}_i \gets H$

            \State $\: \: \: \: \:$ \textbf{Until}  $||\hat{H}^{\pi_j}_i - \hat{H}^{\pi_j}_{i-1} ||< \epsilon$
            
            \State $\: \: \: \: \:$ $H^{\pi_j} \gets \hat{H}^{\pi_j}_i$
        


        \vspace{5pt}
            \State \textbf{Policy improvement:}
            \State $\: \: \: \: \:$ Obtain $w_k^{\pi_{j+1}}    $  such that
            
            \begin{equation}
                \begin{split} 
                    \nonumber
                    \: \: \: \: \: w_k^{\pi_{j+1}}  = - \gamma & (R+ \gamma H_{11}^{\pi_{j}})^{-1} (H_w^{\pi_{j}} w_{k-1,k-n_x+1}   + \\  & H_{\Tilde{y}}^{\pi_{j}} \Tilde{y}_{k,k-n_x+1})             
                \end{split}
            \end{equation} 
       
        \end{algorithmic}
    \end{algorithm}

\section{Simulation results}
\label{sec: simulation}

\begin{figure}[t]
\begin{center}
\includegraphics[width=3in]{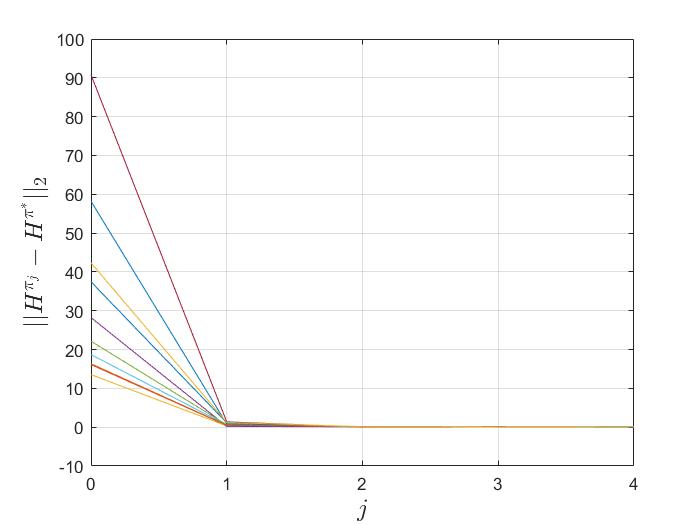}
\caption{Convergence of $H^{\pi_j}$.}
\label{fig: LQE_convergence}
\end{center}
\end{figure}

First, we demonstrate that for a linear system, our proposed method converges to the optimal correction term policy derived from closed-form solution given in the remark.
 Consider   a pendulum modeled as~\eqref{eq: linear model pendulum} with sampling time $T_s=0.1$s.
\begin{equation}
    \label{eq: linear model pendulum}
    \begin{split}
         \begin{bmatrix}
         x_{1,k+1} \\ x_{2,k+1}
         \end{bmatrix} &=\begin{bmatrix}
            \textbf{\: \:}
            0.95  &  0.10 \\
           -0.98  &  0.94
         \end{bmatrix}
         \begin{bmatrix}
        x_{1,k} \\ x_{2,k}
         \end{bmatrix}+
         \begin{bmatrix}
            0.005 \\
            0.098
         \end{bmatrix}u_k   \\
         y_k &= \begin{bmatrix}
            0 & 1 
        \end{bmatrix} 
        \begin{bmatrix}
        x_{1,k} \\ x_{2,k}
         \end{bmatrix}
    \end{split}
\end{equation}
As a result, $M_{\Tilde{y}}$ and $M_w$ are obtained as
\begin{gather}
    \nonumber
    M_w = \begin{bmatrix}
     -1   &     \textbf{\: \:} 0  & -0.95  & -0.92 \\
      \textbf{\: \:}0 &  -1  &   \textbf{\: \:}0.98  &  \textbf{\: \:}0.95 
    \end{bmatrix},  \\
        M_{\Tilde{y}} = \begin{bmatrix}
           -0.82  &  \textbf{\: \:}0.96 \\
            \textbf{\: \:}1.89  & -0.99
    \end{bmatrix} 
\end{gather}
We choose $\gamma = 0.6, \: N=300, \: \beta=0 , \: Q=10$ and $  R= \begin{bmatrix}
    1 & 0 \\ 0 & 1
\end{bmatrix}$. From closed-form solution, we get
\begin{gather}
    P^{\pi^*} = 
\begin{bmatrix} 
   \textbf{\: \:} 1.28  & -0.78 \\
   -0.78  & \textbf{\: \:} 10.77
\end{bmatrix}
\end{gather} 
\begin{multline}
    \label{eq: H_matrix}
    H^{\pi^*} =
    \begin{bmatrix}
   \textbf{\: \:} 1.3 &  -0.8  &   \textbf{\: \:}1.9   &   \textbf{\: \:}1.9    & \textbf{\: \:} 2.5   &  -2.0  \\
   -0.8 &  \textbf{\: \:}10.7   & -11.2   & -10.9   & -21.0   & \textbf{\: \:} 11.4  \\
    \textbf{\: \:}1.9 &  -11.2  &  \textbf{\: \:} 12.9  &  \textbf{\: \:} 12.5  &   \textbf{\: \:}22.9  &  -13.1 \\
    \textbf{\: \:}1.9 &  -10.9  &   \textbf{\: \:}12.5  &  \textbf{\: \:} 12.2 & \textbf{\: \:}  22.3  &  -12.7 \\
    \textbf{\: \:} 2.5  & -21.0   & \textbf{\: \:} 22.9   &  \textbf{\: \:} 22.3   & \textbf{\: \:} 41.8    & -23.2  \\
   -2.0  & \textbf{\: \:} 11.4   & -13.1   & -12.7  & -23.2   & \textbf{\: \:} 13.2  
    \end{bmatrix}
\end{multline}
The objective is to show that  with any initial stabilizing policy $\pi_0(.)$, the proposed algorithm~\ref{alg: PI_LQE_iterative} converges to the matrix $H^{\pi^*}$ and therefore the optimal policy $\pi^*(.)$. Consider ten simulations with random initial stabilizing policies.
It is shown in Fig~\ref{fig: LQE_convergence} that $\pi_j(.)$ converges to the optimal policy $\pi^*(.)$ in all ten runs of the algorithm since $H^{\pi_j}$ converges to $H^{\pi^*}$.

\begin{figure}[t]
\begin{center}
\includegraphics[width=3.3in]{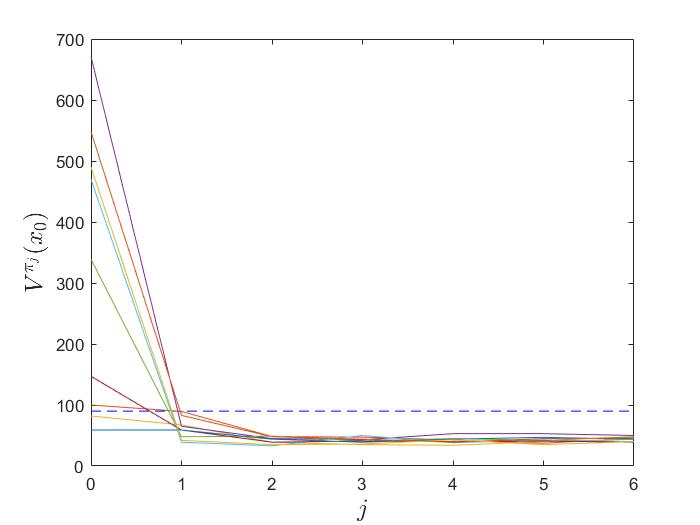}
\caption{Cost-to-go from the initial state \\ over the policy number}
\label{fig: LQE_convergence_nonlinear}
\end{center}
\end{figure}

\begin{figure}[t]
\begin{center}
\includegraphics[width=3.3in]{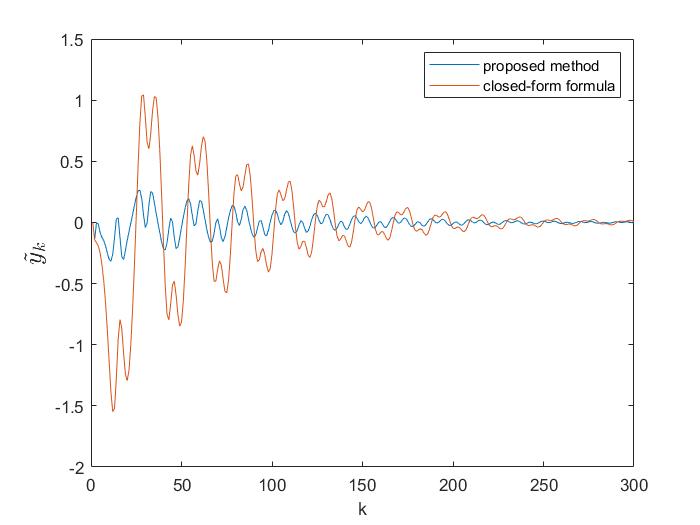}
\caption{Output error comparison}
\label{fig: output_error}
\end{center}
\end{figure}

The proposed method is applied to the nonlinear model of the simple pendulum in the second example. The data-driven method gives  an improved observer compared to  obtaining the closed-form solution of the correction term using the linearized model. 
Consider the pendulum with the dynamics as
    \begin{equation}
         \Ddot{\theta} \left( t \right)  + 0.1 \dot{\theta} \left (t \right)  + 10 \: sin \left (\theta \left (t \right ) \right ) = u(t)
    \end{equation}
    where $\theta(t)$ represents the pitch angle and the measured output is $\dot{\theta} \left (t \right)$.
    We again choose $\gamma = 0.6, \: N=300, \: \beta=0 , \: Q=10$ and $  R= \begin{bmatrix}
    1 & 0 \\ 0 & 1
\end{bmatrix}$.
The $A, \: B, \: C$ matrices in~\eqref{eq: linear model pendulum}
   which are derived from linearization around 
    \begin{gather}
            \begin{bmatrix}
            {\theta} \\ \dot{\theta}
        \end{bmatrix} = \begin{bmatrix}
            0 \\ 0
        \end{bmatrix}
    \end{gather}
    are used to design the observer given in~\eqref{eq: Luenberger observer }. One can also obtain the correction term $w_k$ solely using the linearized model with the closed form solution given in~\eqref{eq: H_matrix}. To compare this result with obtaining the correction term using the proposed data-driven method, we run ten simulation with different initial stabilizing policies and the initial state
        \begin{gather}
            x_0 = 
            \begin{bmatrix}
            {\theta(0)} \\ \dot{\theta}(0)
        \end{bmatrix} = \begin{bmatrix}
            3 \\ 0
        \end{bmatrix}
    \end{gather}
    Figure~\ref{fig: LQE_convergence_nonlinear} shows that the  proposed method consistently yields lower initial cost-to-go values than the closed-form solution, wich is shown with a dash line. 
In this instance, the selection of $Q$ and $R$ matrices prioritizes minimizing the output error. As depicted in Fig~\ref{fig: output_error}, the proposed approach outperforms the correction term derived from the closed-form formula  and reduces the output error over time.

\section{Conclusion}
\label{sec: conclusion}
This paper introduced an innovative policy iteration method to design an optimal observer with a quadratic cost function. This data-driven approach enhances the observer's performance for systems with a linearized model and a stabilizing Luenberger observer gain.
The two-layer quadratic neural network is used as the value-function approximator. The nueral network  has an analytical, quadratic input-output mapping trained with convex optimization.
Therefore a linear correction term policy  is derived from input and output data to rectify inaccuracies of the linearized model. This contrasts with existing techniques using neural networks, which require a second neural network for policy improvement.
 Using the proposed approach for linear systems converges to the optimal observer obtained from analytical methods. Applied to a simple pendulum, our method demonstrates improved correction term policies compared to relying solely on the linearized model, showing its potential for nonlinear systems.

\bibliographystyle{unsrt} 
\bibliography{citation} 

\begin{thebibliography}{10}

\bibitem{luenberger_observer}
David~G Luenberger.
\newblock Observing the state of a linear system.
\newblock {\em IEEE transactions on military electronics}, 8(2):74--80, 1964.

\bibitem{KForg}
R~Kalman.
\newblock A new approach to linear filtering and prediction problems.
\newblock {\em ASME Journal of Basic Engineering}, 82:35--45, 1960.

\bibitem{sutton2018reinforcement}
Richard~S Sutton and Andrew~G Barto.
\newblock {\em Reinforcement learning: An introduction}.
\newblock MIT press, 2018.

\bibitem{lewisoptimalbook}
Frank~L Lewis, Draguna Vrabie, and Vassilis~L Syrmos.
\newblock {\em Optimal control}.
\newblock John Wiley \& Sons, 2012.

\bibitem{busoniu2017reinforcement}
Lucian Busoniu, Robert Babuska, Bart De~Schutter, and Damien Ernst.
\newblock {\em Reinforcement learning and dynamic programming using function
  approximators}.
\newblock CRC press, 2017.

\bibitem{benchmarkingRL}
Yan Duan, Xi~Chen, Rein Houthooft, John Schulman, and Pieter Abbeel.
\newblock Benchmarking deep reinforcement learning for continuous control.
\newblock In {\em International conference on machine learning}, pages
  1329--1338. PMLR, 2016.

\bibitem{bertsekas2019reinforcement}
Dimitri Bertsekas.
\newblock {\em Reinforcement learning and optimal control}.
\newblock Athena Scientific, 2019.

\bibitem{kiumarsi2017optimal}
Bahare Kiumarsi, Kyriakos~G Vamvoudakis, Hamidreza Modares, and Frank~L Lewis.
\newblock Optimal and autonomous control using reinforcement learning: A
  survey.
\newblock {\em IEEE transactions on neural networks and learning systems},
  29(6):2042--2062, 2017.

\bibitem{lewis2009reinforcement}
Frank~L Lewis and Draguna Vrabie.
\newblock Reinforcement learning and adaptive dynamic programming for feedback
  control.
\newblock {\em IEEE circuits and systems magazine}, 9(3):32--50, 2009.

\bibitem{TD2018}
Vitchyr Pong, Shixiang Gu, Murtaza Dalal, and Sergey Levine.
\newblock Temporal difference models: Model-free deep rl for model-based
  control.
\newblock {\em arXiv preprint arXiv:1802.09081}, 2018.

\bibitem{tedrake2016underactuated}
Russ Tedrake.
\newblock Underactuated robotics: Algorithms for walking, running, swimming,
  flying, and manipulation.
\newblock {\em Course Notes for MIT}, 6, 2016.

\bibitem{bertsekas2012dynamic}
Dimitri Bertsekas.
\newblock {\em Dynamic programming and optimal control: Volume I}, volume~4.
\newblock Athena scientific, 2012.

\bibitem{prokhorov1997adaptive}
Danil~V Prokhorov and Donald~C Wunsch.
\newblock Adaptive critic designs.
\newblock {\em IEEE transactions on Neural Networks}, 8(5):997--1007, 1997.

\bibitem{LQR_RL_stable}
Bo~Pang and Zhong-Ping Jiang.
\newblock Robust reinforcement learning: A case study in linear quadratic
  regulation.
\newblock In {\em Proceedings of the AAAI Conference on Artificial
  Intelligence}, volume~35, pages 9303--9311, 2021.

\bibitem{khan2012reinforcement}
Said~G Khan, Guido Herrmann, Frank~L Lewis, Tony Pipe, and Chris Melhuish.
\newblock Reinforcement learning and optimal adaptive control: An overview and
  implementation examples.
\newblock {\em Annual reviews in control}, 36(1):42--59, 2012.

\bibitem{ADHDP}
Dongbin Zhao, Zhongpu Xia, and Ding Wang.
\newblock Model-free optimal control for affine nonlinear systems with
  convergence analysis.
\newblock {\em IEEE Transactions on Automation Science and Engineering},
  12(4):1461--1468, 2014.

\bibitem{QNN}
Burak Bartan and Mert Pilanci.
\newblock Neural spectrahedra and semidefinite lifts: Global convex
  optimization of polynomial activation neural networks in fully
  polynomial-time.
\newblock {\em arXiv preprint arXiv:2101.02429}, 2021.

\bibitem{Luis_QNN}
Luis Rodrigues and Sidney Givigi.
\newblock Analysis and design of quadratic neural networks for regression,
  classification, and lyapunov control of dynamical systems.
\newblock {\em arXiv preprint arXiv:2207.13120}, 2022.

\bibitem{matni2019self}
Nikolai Matni, Alexandre Proutiere, Anders Rantzer, and Stephen Tu.
\newblock From self-tuning regulators to reinforcement learning and back again.
\newblock In {\em 2019 IEEE 58th Conference on Decision and Control (CDC)},
  pages 3724--3740. IEEE, 2019.

\bibitem{bradtke1994adaptive}
Steven~J Bradtke, B~Erik Ydstie, and Andrew~G Barto.
\newblock Adaptive linear quadratic control using policy iteration.
\newblock In {\em Proceedings of 1994 American Control Conference-ACC'94},
  volume~3, pages 3475--3479. IEEE, 1994.

\bibitem{lewis2010reinforcement}
Frank~L Lewis and Kyriakos~G Vamvoudakis.
\newblock Reinforcement learning for partially observable dynamic processes:
  Adaptive dynamic programming using measured output data.
\newblock {\em IEEE Transactions on Systems, Man, and Cybernetics, Part B
  (Cybernetics)}, 41(1):14--25, 2010.

\bibitem{kiumarsi2014reinforcement}
Bahare Kiumarsi, Frank~L Lewis, Hamidreza Modares, Ali Karimpour, and
  Mohammad-Bagher Naghibi-Sistani.
\newblock Reinforcement q-learning for optimal tracking control of linear
  discrete-time systems with unknown dynamics.
\newblock {\em Automatica}, 50(4):1167--1175, 2014.

\bibitem{na2017adaptive}
Jing Na, Guido Herrmann, and Kyriakos~G Vamvoudakis.
\newblock Adaptive optimal observer design via approximate dynamic programming.
\newblock In {\em 2017 American Control Conference (ACC)}, pages 3288--3293.
  IEEE, 2017.

\bibitem{li2020networked}
Jinna Li, Zhenfei Xiao, Ping Li, and Zhengtao Ding.
\newblock Networked controller and observer design of discrete-time systems
  with inaccurate model parameters.
\newblock {\em ISA transactions}, 98:75--86, 2020.

\end{thebibliography}

\end{document}